\documentclass[amsmath,amssymb,aip,jcp,10pt,twocolumn]{revtex4-1}
\usepackage{dcolumn}
\usepackage{graphicx}
\usepackage{setspace}

\begin{document}

\title{Long range interactions between like homonuclear alkali metal diatoms}

\author{Jason N. Byrd}
\email{byrd@phys.uconn.edu}
\author{Robin C\^{o}t\'{e}}
\author{John A. Montgomery, Jr.}
 \affiliation{Department of Physics, University of Connecticut, Storrs, CT 06269}

\begin{abstract}
Long range electrostatic and van der Waals coefficients up to 
terms of order $R^{-8}$ have been evaluated by the sum over states 
method using {\it ab initio} and time dependent density functional 
theory.  We employ several widely used density functionals and 
systematically investigate the convergence of the calculated 
results with basis set size.  Static electric moments and 
polarizabilities up to octopole order are also calculated.  We 
present values for Li${}_2$ through K${}_2$ which are in good 
agreement with existing values, in addition to new results for Rb${}_2$ and
Cs${}_2$.  Interaction potential curves calculated from these results are shown
to agree well with high level {\it ab initio} theory.
\end{abstract}

\maketitle

\section{Introduction}

The long range interactions between atom-diatom and diatom-diatom systems at low
densities and cold temperatures is important in many areas of atomic and molecular physics.
Applications may be found in precision spectroscopy,\cite{demille2008}
condensed matter physics,\cite{micheli2006}
as well as in the study\cite{balakrishnan2001,quemener2009,ni2010} and control\cite{krems2008} of chemical reactions.
In the study of low density diatomic collisions, interactions at long range play
a critical role in the reaction process.\cite{weck2006}
The weakness of the long range intermolecular interaction compared to the chemical bond
and the range of nuclear coordinates necessary to model a long range potential
energy surface (PES) suggests that it is advantageous to consider modeling the
intermolecular potential in other ways than brute force {\it ab initio} quantum chemical calculations.

In the limit where the wavefunction overlap between  two interacting monomers
is negligible, the interaction potential between them can be expanded in a van
der Waals series, where the interaction energy can be broken into three distinct
components
\begin{equation}
E_{\rm int} = E_{\rm el} + E_{\rm ind} + E_{\rm disp}.
\end{equation}
Here $E_{\rm el}$, $E_{\rm ind}$ and $E_{\rm disp}$ are the permanent
electrostatic, induction (permanent-induced electrostatic) and dispersion
energies and can each be written in an asymptotic series 
\begin{equation}
E_{\rm LR} = \sum_{n} V_n R^{-n}.
\end{equation}  
These contributions to the long range interaction energy can be
calculated in several ways. In this work we expand the interaction operator in
terms of multipole moments,\cite{buckingham1967} and then calculate the interaction
energy using first and second order Rayleigh-Schr\"odinger perturbation
theory.

Isotropic dispersion interactions between Na${}_2$ and K${}_2$ pairs have been
previously studied using the London formula\cite{zemke2010} and
time dependent density functional theory (TD-DFT).\cite{banerjee2007,banerjee2009}
Spelsberg {\it et al.}\cite{spelsberg93} have calculated
van der Waals coefficients including anisotropic corrections
using valence full configuration interaction (VFCI) theory for 
Li${}_2$, Na${}_2$ and K${}_2$.
Also recent work by Kotochigova\cite{kotochigova2010} has calculated the
isotropic van der Waals interaction for KRb with leading order anisotropic
corrections.
In this paper we present results for the van der Waals interactions
between pairs of ground state homonuclear alkali metal diatoms,
through order $R^{-8}$, including anisotropic corrections.
In Sec. \ref{vdwsec} an outline of the sum over states method of calculating van der Waals
coefficients is presented, followed by a discussion of the electronic structure
calculations done to obtain the van der Waals coefficients in Sec. \ref{esc}.
We conclude in Sec. \ref{nrd} with a discussion of our results.
The extension to heteronuclear alkali diatomic van der Waals interactions will be
presented in subsequent publications.

\begin{figure}[b]
\caption{\label{alphaplot}Basis set convergence of
the octopole static polarizability values for Na${}_2$ evaluated using
several {\it ab initio} and DFT methods.}
\resizebox{8cm}{!}{\includegraphics{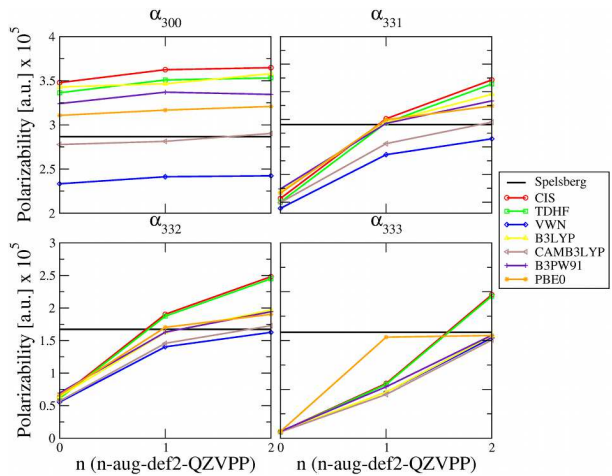}}
\end{figure}

\section{\label{vdwsec}Anisotropic Long-Range Interactions}

\begin{table*}
\begin{ruledtabular}
\begin{tabular}{cddddddd}
$n~L_1~L_2~M$& 
\multicolumn{1}{c}{CIS} &  \multicolumn{1}{c}{TD-HF} &  
\multicolumn{1}{c}{VWN} &  \multicolumn{1}{c}{B3LYP} &  
\multicolumn{1}{c}{CAMB3LYP} &  \multicolumn{1}{c}{B3PW91} &  
\multicolumn{1}{c}{PBE0}\\
\hline
\multicolumn{1}{l}{Dispersion} \\
6   0   0   0 &  8.304(3) &  4.679(3) &  4.094(3) &  3.697(3) &  3.654(3) &  4.216(3) &  4.073(3)\\
6   2   0   0 &  1.192(3) &  5.465(2) &  4.985(2) &  5.408(2) &  5.092(2) &  5.984(2) &  5.487(2)\\
6   2   2   0 &  5.442(2) &  2.128(2) &  2.030(2) &  2.633(2) &  2.363(2) &  2.842(2) &  2.470(2)\\
6   2   2   1 &  -1.209(2) &  -4.728(1) &  -4.511(1) &  -5.851(1) &  -5.251(1) &  -6.315(1) &  -5.489(1)\\
6   2   2   2 &  1.512(1) &  5.910(0) &  5.639(0) &  7.314(0) &  6.564(0) &  7.893(0) &  6.861(0)\\
     \%RMS    &  44.6     &  5.8      &  2.8      &  5.5      &  6.1      &  1.4      &  2.3 \\
\hline                  
\multicolumn{1}{l}{Dispersion} \\
8   0   0   0 &  9.397(5) &  6.569(5) &  5.020(5) &  5.491(5) &  4.923(5) &  6.035(5) &  5.777(5)\\
8   2   0   0 &  4.593(5) &  2.789(5) &  2.261(5) &  3.183(5) &  2.608(5) &  3.324(5) &  3.082(5)\\
8   2   2   0 &  1.242(5) &  6.206(4) &  5.249(4) &  8.974(4) &  6.941(4) &  9.124(4) &  8.025(4)\\
8   2   2   1 &  -1.687(4) &  -8.525(3) &  -7.200(3) &  -1.241(4) &  -9.553(3) &  -1.262(4) &  -1.109(4)\\
8   2   2   2 &  4.759(2) &  3.091(2) &  2.541(2) &  5.072(2) &  3.593(2) &  5.198(2) &  4.481(2)\\
8   4   0   0 &  2.388(4) &  1.116(4) &  9.514(3) &  1.607(4) &  1.340(4) &  1.376(4) &  1.253(4)\\
8   4   2   0 &  1.439(4) &  5.778(3) &  5.170(3) &  1.029(4) &  8.154(3) &  8.760(3) &  7.528(3)\\
8   4   2   1 &  -1.906(3) &  -7.647(2) &  -6.843(2) &  -1.362(3) &  -1.080(3) &  -1.157(3) &  -9.951(2)\\
8   4   2   2 &  9.713(1) &  3.895(1) &  3.484(1) &  6.941(1) &  5.507(1) &  5.862(1) &  5.047(1)\\
\hline                  
\multicolumn{1}{l}{Induction} \\
8   0   0   0 &  1.401(5) &  1.086(5) &  1.073(5) &  4.984(4) &  6.712(4) &  7.579(4) &  7.459(4)\\
8   2   0   0 &  9.006(4) &  6.887(4) &  6.834(4) &  3.237(4) &  4.335(4) &  4.909(4) &  4.802(4)\\
8   2   2   0 &  3.576(4) &  2.425(4) &  2.511(4) &  1.390(4) &  1.784(4) &  2.067(4) &  1.928(4)\\
8   2   2   1 &  -7.153(3) &  -4.849(3) &  -5.022(3) &  -2.779(3) &  -3.569(3) &  -4.134(3) &  -3.855(3)\\
8   2   2   2 &  1.788(3) &  1.212(3) &  1.256(3) &  6.948(2) &  8.922(2) &  1.033(3) &  9.639(2)\\
8   4   0   0 &  6.003(4) &  4.656(4) &  4.598(4) &  2.136(4) &  2.877(4) &  3.248(4) &  3.197(4)\\
8   4   2   0 &  4.721(4) &  3.201(4) &  3.315(4) &  1.834(4) &  2.355(4) &  2.728(4) &  2.545(4)\\
8   4   2   1 &  -6.438(3) &  -4.364(3) &  -4.520(3) &  -2.501(3) &  -3.212(3) &  -3.720(3) &  -3.470(3)\\
8   4   2   2 &  3.576(2) &  2.425(2) &  2.511(2) &  1.390(2) &  1.784(2) &  2.067(2) &  1.928(2)\\
     \%RMS    &  22.1     &  7.3      &  8.0      &  5.5      &  8.6      &  3.0      &  4.3
\end{tabular}
\end{ruledtabular}
\caption{\label{na2vdw} Unique calculated dispersion and induction van
der Waals coefficients, $W^{(2)}_{n L_1 L_2M}$, for ground state Na${}_2$ using
selected {\it ab initio} and DFT
methods.  The RMS deviations are relative to
previous theoretical results.\cite{spelsberg93}  All calculations are performed at
the experimental equilibrium bond length.}
\end{table*}

In the Born-Oppenheimer approximation, the interaction between two linear molecules can be
expanded in a complete angular basis as 
\begin{equation}\label{vdw1}
E_{\rm int} (\hat{r}_1,\hat{r}_2,{\bf R}) = \sum_{L_1 L_2 L} 
E_{L_1 L_2 L}(R) A_{L_1 L_2 L}(\hat{r}_1,\hat{r}_2,\hat{R}),
\end{equation}
where $\hat{r}_i = (\theta_i,\phi_i)$ are the molecular orientations and ${\bf
R}=(R,\theta,\phi)$ defines the vector between the molecular centers. 
As shown by Mulder, van der Avoird, and Wormer,\cite{mulder1979}
if we choose coordinates so that ${\bf R}$ is oriented along the $z$ axis,
the angular functions may be written as
\begin{multline}\label{vdw2}
A_{L_1 L_2 L}(\hat{r}_1,\hat{r}_2,\hat{R}) = \sum_{M=0}^{min(L_1 L_2)} \eta^{M}_{L_1 L_2 L} \\
\times P^{M}_{L_1}(\cos\theta_1) P^{M}_{L_2}(\cos\theta_2)\cos M(\phi_1-\phi_2),
\end{multline}
where
\begin{multline}
\eta^M_{L_1 L_2 L} = (-1)^M (2-\delta_{M,0}) (L_1 M;L_2 -M|L 0) \\
\times \left[\frac{(L_1-M)!  (L_2-M)!} {(L_1+M)!(L_2+M)!}\right]^{1/2},
\end{multline}
$(L_1 M;L_2 -M|L 0)$ is a Clebsch-Gordon coefficient, and $P_L^M(\cos\theta)$
is an associated Legendre polynomial.
Because the interaction energy is rotationally invariant, it may be
expanded in terms of multipole operators
$Q_{lm}=\sum_i z_i r^l_i C_{lm}(\hat{r}_i)$,
where $C_{lm}(\hat{r}_i)$ is a Racah spherical harmonic.\cite{edmonds57}
Using first- and second-order perturbation theory,
Mulder {\it et al.}\cite{mulder1979,avoird1980} express
the coefficients $E_{L_1 L_2 L}(R)$ in terms of the separated
monomer transition moments.
When this is done, the interaction energy may be written as
\begin{multline}\label{vdw3}
E_{\rm int}(R,\theta_1,\phi_1,\theta_2,\phi_2) = \sum_{L_1 L_2}\sum_{M=0}^{min(L_1,L_2)}
V^{(1,2)}_{L_1 L_2 M}(R) \\
P^{M}_{L_1}(\cos\theta_1) P^{M}_{L_2}(\cos\theta_2)\cos M(\phi_1-\phi_2).
\end{multline}
Here the first-order contribution
\begin{equation}\label{v1}
V^{(1)}_{L_1 L_2 M}(R) = W^{(1)}_{n L_1 L_2 M} R^{-n}\delta_{L_1+L_2+1,n}
\end{equation}
\begin{multline}\label{c1}
W^{(1)}_{nL_1L_2M} = (-1)^{L_1+M}(2-\delta_{M,0})\frac{(L_1+L_2)!}{(L_1+M)!(L_2+M)!} \\
  \times \langle 0_1|Q_{L_1 0}|0_1\rangle \langle 0_2|Q_{L_2 0}|0_2\rangle
\end{multline}
is due to the electrostatic interaction and the second-order terms
\begin{equation}\label{v2}
V^{(2)}_{L_1 L_2 M}(R) = -\sum_{l_1  l'_1  l_2  l'_2} C^{l_1  l'_1 L_1;l_2 
l'_2 L_2;M}_n R^{-n} \delta_{l_1+ l'_1+ l_2+ l'_2+2,n}
\end{equation}
\begin{equation}\label{c2}
C^{l_1 l'_1 L_1;l_2 l'_2 L_2;M}_n = \zeta^{l_1 l'_1;l_2 l'_2}_{L_1L_2M} \sum'_{k_1 k_2} 
\frac{T^{0_1 k_1}_{l_1 l'_1 L_1} T^{0_2 k_2}_{l_2 l'_2 L_2}}
{\epsilon_{k_1}-\epsilon_{0_1}+ \epsilon_{k_2}-\epsilon_{0_2}}.
\end{equation}
contain contributions from dispersion and induction.
The $\sum'$ implies that $k_1+k_2\ne 0$ and $\epsilon_{k_i}$ is the energy
of the $k_i$'th state.  
The $\zeta^{l_1 l'_1;l_2 l'_2}_{L_1L_2M}$ coefficient is a scalar coupling term
given by\cite{wormer77}
\begin{multline}\label{zeta}
\zeta^{l_1 l'_1;l_2 l'_2}_{L_1L_2M} = 
(-1)^{l_2+l'_2}
((2L_1+1)!  (2L_2+1)!)^{1/2}\\
\times \left[\frac{ (2l_1+2l_2+1)!  (2l'_1+2l'_2+1)! }
        { (2l_1)!  (2l'_1)!  (2l_2)!  (2l'_2)!}\right]^{1/2}
\sum_L \eta^{M}_{L_1 L_2 L} \\
\times (l_1+l_2 0;l'_1+l'_2 0|L 0)
\begin{Bmatrix}
l_1 & l'_1 & L_1 \\
l_2 & l'_2 & L_2 \\
l_1+l_2 & l'_1+l'_2 & L
\end{Bmatrix},
\end{multline}

\begin{table}[t]
\begin{ruledtabular}
\begin{tabular*}{8,5cm}{ccddd}
System & Method & r_e (a.u.) & \alpha_{\|} (a.u.) & \alpha_{\bot} (a.u.)\\
\hline
Li${}_2$ & B3PW91 & 5.051 & 268.4 & 170.0 \\
         & PBE0 & 5.051 & 262.4 & 168.7 \\
         & Ref. \onlinecite{deiglmayr08} & 5.051 & 305.2 & 162.4 \\
         & Ref. \onlinecite{spelsberg93} & 5.051 & 297.7 & 165.1 \\
\hline
Na${}_2$ & B3PW91 & 5.818 & 352.2 & 206.2 \\
         & PBE0 & 5.818 & 338.7 & 203.7 \\
         & Ref. \onlinecite{deiglmayr08} & 5.818 & 378.5 & 162.4 \\
         & Ref. \onlinecite{spelsberg93} & 5.818 & 370.1 & 201.6 \\
\hline
 K${}_2$ & B3PW91 & 7.416 & 714.6 & 301.5 \\
         & PBE0 & 7.416 & 676.2 & 395.4 \\
         & Ref. \onlinecite{deiglmayr08} & 7.416 & 708.2 & 359.6 \\
         & Ref. \onlinecite{spelsberg93} & 7.379 & 677.8 & 363.1 \\
\hline
Rb${}_2$ & B3PW91 & 7.956 & 830.2 & 453.6 \\
         & PBE0 & 7.956 & 785.8 & 448.3 \\
         & Ref. \onlinecite{deiglmayr08} & 7.956 & 789.7 & 405.5 \\
\hline
Cs${}_2$ & B3PW91 & 8.78  & 1096.0 & 577.8 \\
         & PBE0 & 8.78  & 1032.9 & 568.3 \\
         & Ref. \onlinecite{deiglmayr08} & 8.78  & 1012.2 & 509.0 
\end{tabular*}
\end{ruledtabular}
\caption{\label{dipalpha}Static dipole polarizability values for the $X
{}^1\Sigma^+_g$ homonuclear alkali diatoms calculated at the experimental
equilibrium bond length.}
\end{table}

The coupled transition moment for each monomer is defined as 
\begin{equation}\label{coupledt}
T^{0_i k_i}_{l_i l'_i L_i} = \sum_m \langle 0_i|Q_{l_i m}|k_i\rangle \langle
k_i|Q_{l'_i -m}|0_i\rangle (l_i m;l'_i -m|L_i 0)
\end{equation}
where the indices $k_i$ go over ground and excited states.  These coupled
transition moments also transform the same way as the electrostatic moments
$\langle Q_{L M}\rangle_0$.  It should be noted that from Eq.
\ref{c2}, contributions such as
\begin{equation}\label{c2spec1}
T^{0_1 0_1}_{l_1 l'_1 L_1}\sum_{k_2\ne 0}
\frac{T^{0_2 k_2}_{l_2 l'_2 L_2}}{\epsilon_{k_2}-\epsilon_{0_2}}
+(1\rightleftharpoons 2)
\end{equation}
are associated with $(\mu_1)^2\alpha_2+(1\rightleftharpoons 2)$ type induction
terms.  Contributions to Eq. \ref{c2} such as
\begin{equation}\label{c2spec2}
\sum_{\substack{
k_1\ne 0 \\
k_2\ne 0}}
\frac{T^{0_1 k_1}_{l_1 l'_1 L_1} T^{0_2 k_2}_{l_2 l'_2 L_2}}
{\epsilon_{k_1}-\epsilon_{0_1}+ \epsilon_{k_2}-\epsilon_{0_2}}
\end{equation}
are associated with $\alpha_1\alpha_2$ type dispersion terms and can 
can be related to a Casimir-Polder integral over complex frequencies of coupled
dynamic polarizabilities (see Appendix \ref{casimirappendix}).

It is convenient in practice to collect all terms with the same $R$-dependence
in Eq. \ref{v2} into a single expression
\begin{equation}\label{wdef}
V^{(2)}_{L_1 L_2 M}(R) = -\sum_n \frac{W^{(2)}_{n L_1 L_2 M}}{R^n}.  
\end{equation}
The dispersion and induction contributions to Eq. \ref{wdef} are calculated separately as
\begin{equation}
W^{(2)}_{n L_1 L_2 M} = W^{(2,DIS)}_{n L_1 L_2 M}+W^{(2,IND)}_{n L_1 L_2 M}
\end{equation}
and presented in tables \ref{na2vdw} and \ref{x2vdw}.

\section{\label{esc}Electronic Structure Calculations}

\begin{table}[t]
\begin{ruledtabular}
\begin{tabular*}{8.5cm}{l@{}d@{}d@{}d}

        & \multicolumn{1}{c}{Li${}_2$} & \multicolumn{1}{c}{Na${}_2$} & \multicolumn{1}{c}{K${}_2$} \\
\hline
B3PW91                  & 2.649(3) & 4.216(3)  & 1.242(4) \\
PBE0                    & 2.593(3) & 4.073(3)  & 1.185(4) \\
SAOP                    &          & 4.460(3)\footnote{Ref.  \onlinecite{banerjee2007}}
                                                      & 1.106(4)\footnote{Ref.
                                                        \onlinecite{banerjee2009}} \\
London\footnote{Ref. \onlinecite{zemke2010}}          &          & 4.374(3)  & 1.186(4) \\
Valence FCI\footnote{Ref. \onlinecite{spelsberg93}}  & 2.730(3) & 4.181(3)  & 1.039(4) \\

\end{tabular*}
\end{ruledtabular}
\caption{\label{c6iso}Comparison of theoretical isotropic $V_6$ van der Waals
coefficients.}
\end{table}

\begin{table*}
\begin{ruledtabular}
\begin{tabular}{lddddd}
$n~L_1~L_2~M$  & \multicolumn{1}{c}{Li${}_2$} & \multicolumn{1}{c}{Na${}_2$} & \multicolumn{1}{c}{K${}_2$} 
  & \multicolumn{1}{c}{Rb${}_2$} & \multicolumn{1}{c}{Cs${}_2$} \\
\hline
$Q_{20}$  & 10.74    & 10.52    & 15.68    & 16.06    & 27.85    \\
$5 2 2 0$ & 6.921(2) & 6.640(2) & 1.475(3) & 1.548(3) & 1.912(2) \\
$5 2 2 1$ &-1.538(2) &-1.476(2) &-3.278(2) &-3.439(2) &-4.248(2) \\
$5 2 2 2$ & 9.612(0) & 9.223(0) & 2.049(1) & 2.149(1) & 2.655(1) 
\end{tabular}
\end{ruledtabular}
\caption{\label{qq} Electrostatic van der Waals coefficients,
$W^{(1)}_{n L_1 L_2M}$, for the ground state alkali dimers Li${}_2$,
Na${}_2$, K${}_2$, Rb${}_2$ and Cs${}_2$ calculated at the
CCSD(T) level of theory using the finite field method.}
\end{table*}

Transition moment calculations for use in the evaluation of Eq.
\ref{coupledt} were done for the homonuclear alkali metal diatoms
Li${}_2$, Na$_2$, K$_2$, Rb$_2$, and Cs${}_2$
in the $X {}^1\Sigma^+_g$ ground state using a locally modified version
of the GAMESS\cite{gamess1993,gamess2005} quantum chemistry program package.
All calculations have been performed at the experimental
equilibrium bond distances.
The number of excited states included in the sum
in Eq. \ref{c2} is taken to be the total number of
virtual orbitals present.
Tests were performed to ensure convergence of the calculated
TD-DFT transition moments with respect to the grid size.
The grid employed in production calculations uses
155 radial points for all atoms, and prunes
from a Lebedev grid whose largest size is 974,
thus using about 71,000 grid points/atom (the JANS=2 grid in GAMESS).

To provide consistent results for all the alkali metals, the Karlsruhe
def2-QZVPP\cite{weigend2003} basis sets were chosen for this work. 
The def2 basis sets are available for almost the
entire Periodic Table and are well-known for their robustness
and their excellent cost-to-performance ratio in large-scale
Hartree-Fock and density functional theory (DFT) calculations.
The def2-QZVPP basis sets for the alkali metals
contain $spdf$ basis functions with two polarization functions.
For the row five and lower atoms, the inner-core electrons
are replaced by an effective core potential (ECP) to
reduce the number of electrons included in the correlation treatment and to
account for scalar-relativistic effects.  The ECP-28\cite{leininger1996} and
ECP-46\cite{leininger1996} Stuttgart pseudopotentials were used for Rb and Cs respectively.
These ECP definitions leave the sub-valence $s$ and $p$ electrons free, which
are known to contribute the most to core-valence correlation
energy\cite{iron2003} in alkaline systems.  In the transition moment
calculations presented here, core-valence correlation is accounted for
implicitly within the DFT formalism.

When examining the effects of basis set convergence of the van der Waals coefficients it
is convenient to first look at the convergence of the multipole polarizability
defined as 
\begin{equation}
\alpha_{ll'm}=2\sum'_k 
\frac{\langle 0|Q_{l m}|k\rangle\langle k|Q_{l' -m}|0\rangle}
{\epsilon_k-\epsilon_0}.
\end{equation}
The octopole polarizability terms $\alpha_{300}$ are the most sensitive to the
effects of higher angular momentum functions in the basis set, and were therefore
used to asses the convergence of our results with respect to basis set size. 
Test calculations including a sequence of
$g$ basis functions ranging from valence to diffuse indicated
that there was no need to add higher angular
momentum functions to the def2 basis sets.
The effects of adding
additional sets of diffuse even tempered $spdf$ basis functions
can be seen in Fig. \ref{alphaplot}. 
It was found that two additional sets of diffuse functions
was necessary to achieve accurate results for our test case of Na${}_2$.
For a subset of the DFT functionals considered here, the addition of a third set of
diffuse functions was investigated and found to contribute little to the
calculated polarizabilities.
The def2-QZVPP basis sets for K, Rb and Cs include an
additional $spdf$ diffuse function by definition and so only required
a single additional set of diffuse basis functions.
We note that the dipole polarizability, and hence the $V_6$ van der Waals coefficients,
are essentially converged using the unmodified def2 basis set.
A similar procedure was recently used by Rappoport and Furche\cite{furche2010}
to optimize the Karlsruhe def2 basis sets for molecular property calculations.


Using the double augmented basis set (d-aug-def2-QZVPP) described above  
two {\it ab initio} methods were considered for our test case Na${}_2$,
configuration interaction singles\cite{foresman1992,dreuw2005} (CIS) and time
dependent Hartree-Fock\cite{tawada2004,dreuw2005} (TD-HF) theory.  Additionally
we used time dependent density functional theory
(TD-DFT)\cite{tawada2004,dreuw2005} to evaluate excitation energies and
multipole transition moments.  The results of the {\it ab initio} and TD-DFT
calculations are compared to those of
Spelsberg and Meyer\cite{spelsberg93} in Table \ref{na2vdw}.
Despite the success\cite{mulder1979} in describing the van der Waals
interaction of H${}_2$, CIS significantly over estimates the dispersion and
induction van der Waals interaction.  This is not unexpected as CIS is
known not to obey the Thomas-Reiche-Kuhn dipole sum
rule,\cite{thomas1925,reiche1925,kuhn1925}
because it overestimates the transition moments.
An additional source of error in CIS is the overestimation of
excitation energies that results from the use of canonical Hartree-Fock
virtual orbital energies.
The overall performance of TD-DFT
compared to the results of Spelsberg and Meyer\cite{spelsberg93} appears quite
good.  Upon examining the results of Table \ref{na2vdw} we have chosen the
B3PW91 and PBE0 functionals for use in the rest of this work due to the
consistent accuracy for both dispersion and induction interactions.

The electric quadrupole moment used to calculate V${}_5$ was evaluated using the
finite field method using CCSD(T).  Core-valence contributions are accounted
for by including the inner valence $s$ and $p$ electrons in the correlation
treatment.  For consistency the n-def2-QZVPP basis sets defined above were used.
Comparisons with the results of Harrison and Lawson\cite{harrison2005} for
Li${}_2$, Na${}_2$ and K${}_2$ show that this choice of basis set provides
reliable results.  In Table \ref{qq} we show our calculated quadrupole moment
and $V_5$ van der Waals coefficients for the homonuclear alkali diatoms
including new results for Rb${}_2$ and Cs${}_2$.

\section{\label{nrd}Numerical Results and Discussion}

\begin{turnpage}
\begin{table*}
\begin{ruledtabular}
\begin{tabular*}{17cm}{c@{}d@{}d|@{}d@{}d|@{}d@{}d|@{}d@{}d}
             & \multicolumn{2}{c}{Li${}_2$} & \multicolumn{2}{c}{K${}_2$}
             & \multicolumn{2}{c}{Rb${}_2$} & \multicolumn{2}{c}{Cs${}_2$}\\
$n~L_1~L_2~M$& \multicolumn{1}{c}{B3PW91} & \multicolumn{1}{c}{PBE0} 
             & \multicolumn{1}{c}{B3PW91} & \multicolumn{1}{c}{PBE0} 
             & \multicolumn{1}{c}{B3PW91} & \multicolumn{1}{c}{PBE0} 
             & \multicolumn{1}{c}{B3PW91} & \multicolumn{1}{c}{PBE0} \\
\hline
\multicolumn{2}{l}{Dispersion} \\
6 0 0 0& 2.649(3) &  2.593(3) &   1.242(4) &  1.185(4) &  1.600(4) &  1.530(4) & 2.457(4) &  2.334(4) \\
6 2 0 0& 2.996(2) &  2.836(2) &   1.939(3) &  1.722(3) &  2.652(3) &  2.355(3) & 4.428(3) &  3.925(3) \\
6 2 2 0& 1.187(2) &  1.086(2) &   9.994(2) &  8.237(2) &  1.442(3) &  1.186(3) & 2.583(3) &  2.133(3) \\
6 2 2 1&-2.638(1) & -2.414(1) &  -2.221(2) & -1.830(2) & -3.205(2) &  -2.636(2)& -5.739(2)&  -4.740(2)\\
6 2 2 2& 3.297(0) &  3.017(0) &   2.776(1) &  2.288(1) &  4.006(1) &  3.295(1) & 7.174(1) &  5.925(1) \\
\%RMS  & 2.7      &  3.4      &   8.8      &  6.4      &           &           &          &           \\
\hline
\multicolumn{2}{l}{Dispersion} \\
8 0 0 0& 3.170(5) &  3.075(5)   &  2.714(6) &  2.553(6) &  3.928(6) &  3.715(6) & 7.249(6)&  6.792(6) \\
8 2 0 0& 1.302(5) &  1.236(5)   &  1.714(6) &  1.553(6) &  2.615(6) &  2.374(6) & 5.288(6)&  4.762(6) \\
8 2 2 0& 2.963(4) &  2.717(4) &  5.140(5) &  4.333(5) &  8.191(5) &  6.897(5) & 1.788(6)&  1.499(6) \\
8 2 2 1&-4.118(3) & -3.771(3)   & -7.152(4) & -6.032(4) &  -1.132(5) &  -9.527(4) & -2.466(5)&  -2.068(5) \\
8 2 2 2& 1.824(2) &  1.629(2)  &  3.217(3) &  2.735(3) &  4.558(3) &  3.811(3) & 9.693(3)&  8.080(3) \\
8 4 0 0& 1.592(3) &  1.698(3) &  9.509(4) &  8.304(4) &  1.801(5) &  1.629(5) & 4.709(5)&  4.216(5) \\
8 4 2 0& 1.318(3) &  1.284(3) &  6.279(4) &  5.068(4) &  1.247(5) &  1.045(5) & 3.512(5)&  2.933(5) \\
8 4 2 1&-1.721(2) & -1.680(2) & -8.307(3) & -6.706(3) &  -1.653(4) &  -1.386(4) & -4.665(4)&  -3.896(4) \\
8 4 2 2& 8.383(0) &  8.254(0) &  4.225(2) &  3.413(2) &  8.456(2) &  7.101(2) & 2.402(3)&  2.007(3) \\
\hline
\multicolumn{2}{l}{Induction} \\
8 0 0 0& 6.032(4) &  3.075(5)  &  3.447(5) &  3.390(5) &  4.223(5) &  4.199(5) & 6.083(5)&  6.174(5) \\
8 2 0 0& 3.837(4) &  1.236(5)  &  2.254(5) &  2.196(5) &  2.779(5) &  2.736(5) &  4.036(5)&  4.057(5) \\
8 2 2 0& 1.393(4) &  2.717(4) &  1.016(5) &  9.269(4) &  1.308(5) &  1.203(5) & 2.000(5)&  1.889(5) \\
8 2 2 1&-2.787(3) & -3.771(3)  & -2.032(4) & -1.854(4) &  -2.615(4) &  -2.407(4) & -4.000(4)&  -3.778(4) \\
8 2 2 2& 6.967(2) &  1.629(2)  &  5.080(3) &  4.634(3) &  6.538(3) &  6.016(3) & 1.000(4)&  9.445(3) \\
8 4 0 0& 2.585(4) &  1.698(3)  &  1.477(5) &  1.453(5) &  1.810(5) &  1.800(5) & 2.607(5)&  2.646(5) \\
8 4 2 0& 1.839(4) &  1.284(3)  &  1.341(5) &  1.224(5) &  1.726(5) &  1.588(5) & 2.640(5)&  2.494(5) \\
8 4 2 1&-2.508(3) & -1.680(2) & -1.829(4) & -1.668(4) &  -2.354(4) &  -2.166(4) & -3.600(4)&  -3.400(4) \\
8 4 2 2& 1.393(2) &  8.254(0) &  1.016(3) &  9.269(2) &  1.308(3) &  1.203(3) &  2.000(3)&  1.889(3)\\
\%RMS  & 4.5 & 5.4 & 7.2 &  5.9   
\end{tabular*}
\end{ruledtabular}
\caption{\label{x2vdw} Unique van der Waals coefficients,
$W^{(2)}_{n L_1 L_2M}$, for the ground state alkali dimers
Li${}_2$, K${}_2$ and new results for Rb${}_2$ and Cs${}_2$ calculated at the
TD-DFT level of theory using the B3PW91 and PBE0 functionals.  The RMS
deviations are relative to previous theoretical
results\cite{spelsberg93} where applicable.}
\end{table*}
\end{turnpage}

As noted above, the Karlsruhe def2 basis sets replace the inner shell electrons
of rubidium and cesium with an effective core potential.  In order to assess the
effect of replacing the inner-core electrons in heavy atoms by a pseudopotential,
we have performed calculations of the K${}_2$ van der Waals coefficients using
the Stuttgart ECP and basis set,\cite{leininger1996}
and made comparisons with our all-electron calculations using the def2 basis set.
The Stuttgart $sp$ ECP basis set was
uncontracted, and to ensure that the basis set was fully saturated the most
diffuse five $p$ exponents were added as $d$ functions and the subsequent most
diffuse four $d$ exponents were added as $f$ functions. 
Using this basis and ECP, it was found that the difference between the ECP
and all-electron def2 results were negligible.

An important consideration for any van der Waals study is that of the dipole
polarizability of the individual molecules as the leading dispersion interaction
term $V_6$ is proportional to the product of polarizabilities.  Using our
n-aug-def2-QZVPP basis sets with the B3PW91 and PBE0 DFT functionals we have
calculated the dipole, quadrupole and octopole polarizabilities for all the
alkali diatoms considered in this work.  In Table \ref{dipalpha} the dipole
polarizabilities are presented with several other existing theoretical results in
comparison.  The calculated higher order polarizabilities can be found in the
supplemental material.\cite{jcpsupp}
Because the polarizability is very sensitive to the diatomic bond distance
near equilibrium,\cite{deiglmayr08}
we have chosen to perform all calculations at the experimental
equilibrium bond length, rather than at the optimized geometry for each different
level of theory.  As a consequence, there will be
small deviations in the parallel components of the calculated polarizability
at a particular level of theory
compared to what would be found at the optimized geometry.
Therefore, our calculated values for the perpendicular dipole polarizability
($\alpha_{111}$) agrees better with the existing theoretical results than the
parallel polarizability ($\alpha_{110}$).  
For the higher order polarizabilities, the only
available values come from Spelsberg {\it et al.}\cite{spelsberg93}.  The
total polarizability RMS error for Li${}_2$, Na${}_2$ and K${}_2$ is 8.7\%,
6.7\% and 9.5\% and 7.5\%, 5.1\% and 6.9\% for B3PW91 and PBE0 respectively.

\begin{figure}
\resizebox{7.5cm}{!}{\includegraphics{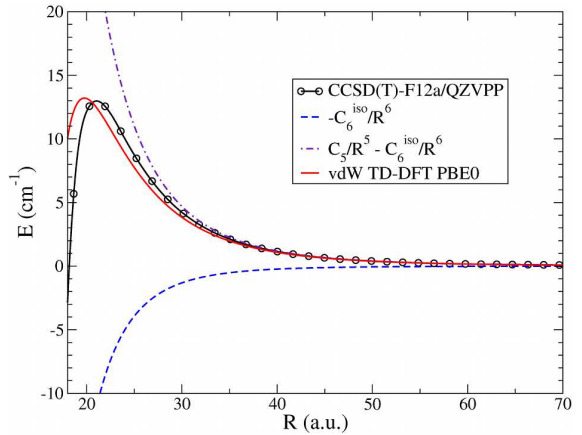}}
\caption{\label{na2comp} Comparison of the van der Waals surface of co-linear
Na${}_2$ with both the London isotropic $V_6$ approximation\cite{zemke2010} and
a fully {\it ab initio} curve.}
\end{figure}

Often neglected, the leading order term in the van der Waals expansion for
homonuclear alkali diatoms is the electrostatic $V_5$ quadrupole-quadrupole
interaction.  This leads to a repulsive interaction potential for a significant
portion of the interaction phase space with a long range barrier; for the case
of Na${}_2$ this barrier has a maximum height of $\sim13$cm${}^{-1}$ for the
co-linear orientation ($\theta_1=\theta_2=0$).  There are however orientations
of approaching pairs of molecules for which no barriers are found.
For example, this was seen for Na${}_2$ and K${}_2$ by
Zemke {\it et al}\cite{zemke2010} as well as for the related systems of
K${}_2+$Rb${}_2$\cite{byrd2010} and Rb${}_2+$Cs${}_2$\cite{tscherbul2008} where the
lowest energy reaction path was calculated {\it ab initio} and indeed was found
to be barrier less.  To illustrate the effect of including the $V_5$
electrostatic term as well as the higher order $V_8$ terms, the results of the
London approximation given by Zemke {\it et al.}\cite{zemke2010} are compared in
Fig. \ref{na2comp} to
the TD-DFT/PBE0 induction and dispersion coefficients reported above for the
co-linear orientation.  As a comparison, the potential curve was also calculated using
the CCSD(T)-F12a (explicitly correlated CCSD(T)) level of
theory\cite{adler2007,knizia2009}.  As can be seen in Fig. \ref{na2comp}, the
inclusion of $V_5$ greatly improves the long range representation of the
potential, while the higher order van der Waals terms are necessary to describe
the interaction as the molecule separation decreases.  With the inclusion of the
higher order terms reported in this work, the van der Waals surface matches the
barrier height and molecule separation to a few percent.

In a cold ($J=0$) ensemble with no external field the rotationally
averaged electrostatic $V_5$ van der Waals coefficient evaluates to zero.  For
this case the leading order term in the van der Waals series is the isotropic
$V_6$ dispersion interaction.  We compare our calculated $V_6$ coefficients to
existing theoretical results in Table \ref{c6iso}.  For the case of Li${}_2$ it
can be seen that B3PW91 performs well, which is also true from Table
\ref{na2vdw} for the case of Na${}_2$.  For the heavier K${}_2$ we find that
PBE0 performs marginally better for $V_6$.  In Table \ref{x2vdw} all the
calculated van der Waals coefficients including all anisotropic contributions
for Li${}_2$ and K${}_2$ are presented.  To compare with the results of
Spelsberg {\it et al.}\cite{spelsberg93} it is necessary to recouple their
interaction coefficients from the $L_1,L_2,L$ coupling scheme to the $L_1,L_2,M$
coupling scheme used here by the following
\begin{equation}
V_{L_1 L_2 M} = \sum_L \eta^M_{L_1 L_2 L} V_{L_1 L_2 L}.
\end{equation}
After recoupling we find our van der Waals coefficients to be in good agreement
with those of Spelsberg {\it et al.}.  Also in Table \ref{x2vdw} are our new
results for Rb${}_2$ and Cs${}_2$.  Using the aug-def2-QZVPP basis set and ECP,
our calculations for
Rb${}_2$ and Cs${}_2$ contain comparable numbers of electrons to the all-electron
calculations for Na$_2$, and therefore
we expect our van der Waals coefficients for these alkali diatoms to
be comparably accurate.

\section{Conclusion}

In summary, we have used {\it ab initio} and TD-DFT methods to
calculate excitation energies and transition moments for the homonuclear
alkali diatoms Li$_2$, Na$_2$, K$_2$, and Cs${}_2$. 
These energies and moments were used in a sum over
states calculation of the electrostatic, induction and dispersion van der
Waals coefficients as well as static polarizabilities up to third order.
The Karlsruhe def2-QZVPP basis sets were augmented with diffuse functions
to ensure accurate calculation of the multipole polarizabilities and
van der Waals coefficients.
Calculated dipole polarizabilities were found to be consistent with other
theoretical results, while noting that the polarizability is sensitive to
the diatomic bond length.\cite{deiglmayr08}
Van der Waals results for the lighter systems are in good agreement with the existing
literature, and suggest errors of a few percent for the new results for
Rb${}_2$ and Cs${}_2$.  As such the long range interaction of each alkali
diatomic pair can be characterized by a van der Waals series of the form 
\begin{multline}\label{simple}
E_{\rm int}(R,\theta_1,\phi_1,\theta_2,\phi_2) = 
\sum_{n L_1 L_2 M} \frac{(W^{(1)}_{n L_1 L_2 M}-W^{(2)}_{n L_1 L_2 M})}{R^n} \\ 
\times P^{M}_{L_1}(\cos(\theta_1)) P^{M}_{L_2}(\cos(\theta_2))\cos(M(\phi_1-\phi_2))
\end{multline}
where the $W^{(1,2)}_{L_1 L_2 M}$ coefficients can be taken directly from Tables
\ref{na2vdw} and \ref{x2vdw}.  We provide a sample FORTRAN program for
evaluating Eq. \ref{simple} in the supplimental material for the case of
Na${}_2$ as described in Fig. \ref{na2comp}.
The current study suggests that the calculation of the van der Waals expansion
coefficients can provide an accurate model of the long range potential surface
in alkali dimers.

It is noted that the leading order van der Waals coefficient is the
electrostatic $V_5$ quadrupole-quadrupole interaction.  This term provides a
molecular frame repulsive interaction at long range for a portion of the
diatom-diatom phase space.  This repulsive interaction will lead to a
stabilization of the molecules when confined in a trap.  Coupling with an
external field can lead to a richer interaction phase space beyond that of the
isotropic dispersion interaction.  

\section{Acknowledgments}

J.N.B would like to thank the US Department of Energy Office of Basic Energy
Sciences for support and R.C. thanks the Department of Defense Air Force Office
of Scientific Research for partial support.

\appendix

\section{Relationship of the sum over states method to the Casimir-Polder
formulation}\label{casimirappendix}

The sum over states formulation of Eq. \ref{c2} is not the only way to
obtain dispersion interactions as described in Eq. \ref{c2spec2}.  We
define the coupled dynamic polarizability as 
\begin{equation}
\alpha^{l_i l'_i}_{L_i}(\omega) = \sum_{k_i\ne 0} 
\frac{\Delta\epsilon_{k_i}T^{0_i k_i}_{l_i l'_i L_i}}
{\Delta\epsilon^2_{k_i} - \omega^2}
\end{equation}
where $\Delta\epsilon_{k_i}=\epsilon_{k_i}-\epsilon_{0_i}$ is the $k_i$'th
excitation energy.  Using the identity
\begin{equation}
\frac{1}{a+b} = \frac{2}{\pi}\int^\infty_0 
\frac{ab}{(a^2+\omega^2)(b^2+\omega^2)}
d\omega 
\end{equation}
then Eq. \ref{c2spec2} can be rewritten as
\begin{equation}
\frac{2}{\pi}\int^\infty_0 
\alpha^{l_1 l'_1}_{L_1}(i\omega) \alpha^{l_2 l'_2}_{L_2}(i\omega)
d\omega 
\end{equation}
which is the well known Casimir-Polder integral for dispersion interactions
between two molecules.

%

\end{document}